# Robust Diffusive Proton Motions in Phase IV of Solid Hydrogen

Hanyu Liu,[1,2] John Tse,[1,2] Yanming Ma[1],*

[1]*State Key Laboratory of Superhard Materials, Jilin University, Changchun 130012, China*

[2]*Department of Physics and Engineering Physics, University of Saskatchewan, Saskatchewan, Canada, S7N 5E2*

**Abstract**

Systematic first-principles molecular dynamics (MD) simulations with long simulation times (7-13 ps) for phase IV of solid hydrogen using different supercell sizes of 96, 288, 576, and 768 atoms established that the diffusive proton motions process in the graphene-like layer is an intrinsic property and independent of the simulation cell sizes. The present study highlights an often overlook issue in first-principles calculations that long time MD is essential to achieve ergodicity, which is mandatory for a proper description of dynamics of a system. The present results contradict a recent work [Phys. Rev. B 87, 174110 (2013)] in which the analysis was relied on short time slices (1-3 ps).

PACS numbers: 66.30.-h; 61.50.Ks; 62.50.-p; 67.80.F-

# Introduction

Solid hydrogen is one of the central topics of condensed matter physics.[1, 2] Recent advances in experimental technique have led to the discovery of a new phase IV at room temperature.[3-8] However, the structural and dynamical properties of this phase IV are under extensive debate.[4, 9, 10] Initially, phase IV was interpreted as having an orthorhombic *Pbcn* symmetry (24 molecules per cell),[11] which consists of alternate layers of graphene-like, three-molecule rings with elongated $H_2$ molecules, and unbound $H_2$ molecules.[4, 11] Theoretical phonon calculations, however, shows this structure is dynamically unstable. A monoclinic *Pc* structure that doubles that of the *Pbcn* structure with 96 atoms per cell and does not contain any imaginary frequency was proposed later.[12] In the meanwhile, a first-principles metadynamics study[13] proposed an alternate *Cc* structure containing 96 molecules per unit cell, geometrically rather similar to *Pc* structure. It was further suggested[13] that phase IV has a partially disordered structure: disorder in the weakly coupled or unbounded $H_2$ molecules layers but ordered in the strongly coupled graphene-like layers (G-layer).

More recently, a novel diffusive hydrogen process with collaborative rotation of three hydrogen molecules in a ring fashion in the graphene-like layers at a pressure range of 250–350 GPa and temperature range between 300–500 K was reported.[7, 14] The finding of the proton diffusion provides an explanation on the observed abrupt increase of Raman linewidth at the formation of phase IV and its large increase with pressure.[7, 14] Other *ab initio* MD simulations[15, 16] on phase IV found similar results.[7, 14] These studies show (i) the trend of the experimental Raman spectrum can be explained by the *Pc* or *Cc* structures;[16] (ii) at 250 GPa and 220 K, a simulation started from phase III (*C2/c*) was found to transform into phase IV by heating the system to 300 K;[15] (iii) rotation of the 3-molecules ring in the G-layer of phase IV predicted in Ref. 7 and 14 was confirmed by another MD simulations.[15] In Ref. 15, the authors however claimed that there are more than one G-layer in phase IV (G' and G'') and diffusive hydrogen motions do not exist. They further argued that these discrepancies with other MD calculations[7, 14] were attributed to finite size effects.[15] It is noteworthy that a very recent *ab initio* MD calculation[17] also reported atomic diffusion in a molecular solid hydrogen phase (*Cmca*-12) when temperature goes beyond 429 K in the pressure range of 260-350 GPa. Moreover, the hydrogen transfer was found to confine within one type of layer,[17] similar to the diffusive hydrogen motions observed in the G-layer of phase IV.[7, 14]

A major difference between Ref. 15 and the other studies[7, 14] is the length of the simulation time. In Refs. 7 and 14, MD simulation times are ranging from 7 ps[14] to 90 ps[7]. This is in clear contrast to the simulation time span of only 1-3 ps in Refs. 15 and 16. It is well established that short-time MD simulation does not sample the potential landscape adequately and might not capture important events. In this study, first-principles MD simulations with much longer simulation times (7-13 ps) at 300 GPa and 300 K for phase IV of solid hydrogen using different supercell sizes of 96, 288, 576, and 768 atoms are systematically performed to examine the finite size effects. Through carefully analysis of the atomic trajectories, it is found that the above discrepancies on the description of the hydrogen dynamics in the G-layers are not due to cell sizes but rather on the time span of the MD calculations. Rationally, results derived from short time MD simulations that are not long enough to satisfy the ergodicity assumption can be unreliable. Our current results confirmed previous MD results[7, 14] and allows us to conclude that the diffusive proton motions in G-layers for phase IV are robust.

**Computational details**

*Ab initio* MD simulations for solid hydrogen were performed at 300 GPa and 300 K with the NPT[18] (*N*-number of particles, *P*-pressure, *T*-temperature) ensemble recently implemented in *Vienna Ab initio Simulation Package* (VASP) code.[19-21] The all-electron projector-augmented wave (PAW) method was adopted.[22] Exchange and correlation effects were treated in generalized-gradient approximation (GGA).[23] We have adopted simulation models containing 96, 288, 576 and 768 atoms for phase IV with simulation times ranging from 7 to 13 ps. A plane-wave cutoff of 500 eV was used. Brillouin zone sampling of 4×4×4, 2×2×4, 2×2×2, 2×2×2 $k$-meshes[24] for supercells consisting of 96 (*Cc*), 288 (*Pc*), 576 (*Cc*) and 768 (*Cc*) atoms, respectively, was employed. The computational parameters chosen were sufficient to produce converged results. The time step used in the MD simulation was 0.5 fs and the self-consistency on the total energy was chosen to be $1\times10^{-5}$ eV.

## Results and discussions

The atomic mean square displacement (MSD) is an important method to investigate the diffusion of a system. We computed the MSD of hydrogen for phase IV as derived from the present MD simulations at 300 GPa and 300 K for different supercell sizes.

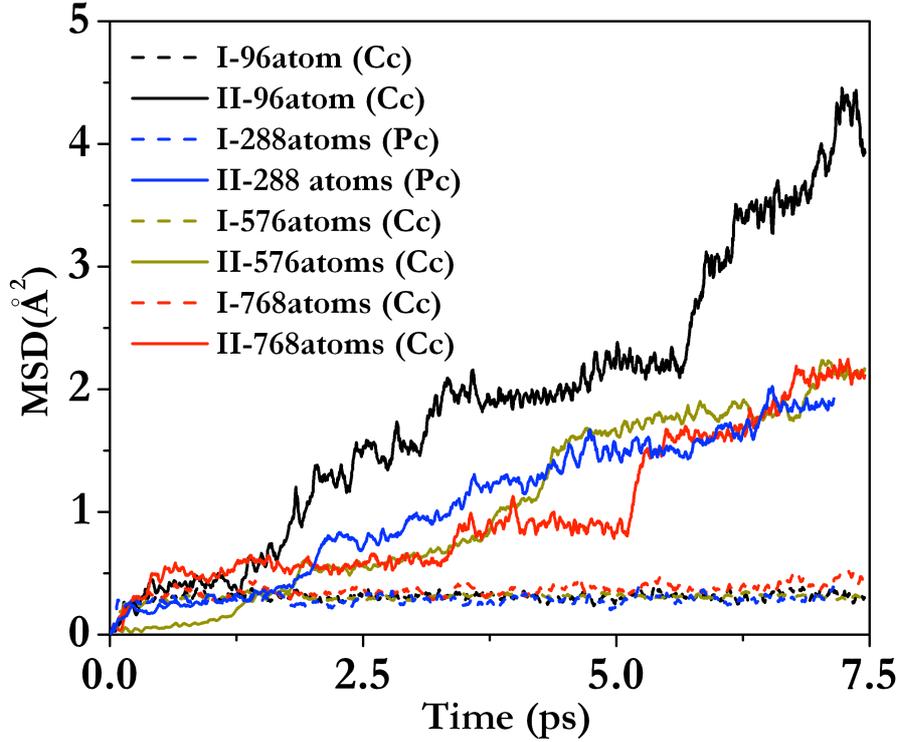

**Fig. 1**: Mean squared displacements as a function of time at 300 GPa and room temperature for phase IV of solid hydrogen with different simulated supercells of 96, 288, 576 and 768 atoms are shown, respectively. Layer I and II is a freely rotating hydrogen molecules layer and strongly coupled graphene-like layer (G-layer), respectively.

The results summarized in Fig. 1 show unambiguously that the hydrogen MSD in the G-layer (or layer II) grows with time, even in the largest supercell (768 atoms). This observation demonstrates unequivocally the proton diffusion at 300 GPa and 300 K.

There is a weak dependence on the MSD values with the size of the simulation cells, showing that the MSD decreases with increasing cell size at smaller size regime. However, this value is well converged when the cell size reaches 288 atoms (Fig. 1). From the slopes of the MSD curves, the averaged diffusion coefficients are estimated to be ~$0.7 \times 10^{-5}$ cm$^2$/s for supercells of 288, 576 and 768 atoms. This diffusion magnitude is comparable to that for a typical liquid, e.g., ~$2.3 \times 10^{-5}$ cm$^2$/s in liquid water at 298.16 K.[25] In this manner, hydrogen in the G-layers may be considered as a sub-lattice melting. However, unlike water, hydrogen diffusion in phase IV of the solid is not a random process but follows a well-defined pattern through a series of hopping motions.

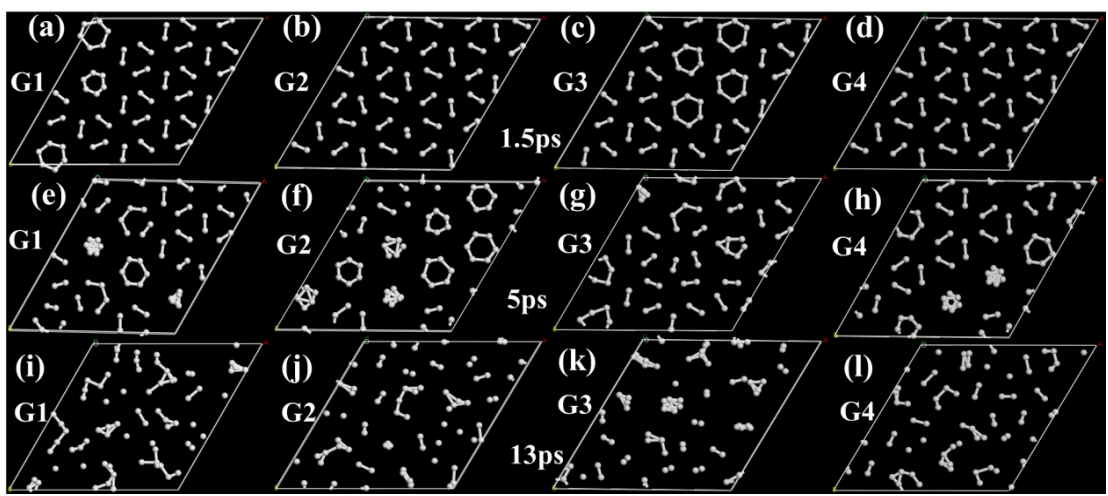

**Fig. 2**: Averaged atomic positions of different G-layers for phase IV using supercells of 576 atoms at 300 GPa and 300 K from 1.5 ps (top panel), 5 ps (middle panel) and 13 ps (bottom panel) *NPT*-MD simulation, respectively.

From the analysis of the averaged atomic positions of the *Pc* structure,[15] Magdau and Ackland proposed that phase IV of solid hydrogen adopts a dynamic, simple hexagonal structure with three layers: a freely rotating hydrogen molecules layer, a static hexagonal trimers layer, and a rotating hexagonal trimers layer. It is correct to interpret the crystal structure with the averaged atomic positions if the atoms are not diffusive. However, it is not applicable in phase IV. As already shown above, the hydrogen atoms in G-layer diffuse substantially. Therefore, characterization of this "liquid-like" structure by averaging the atomic positions may not be appropriate. To demonstrate such a failure, we have computed the averaged atomic positions of phase IV for different running time durations at 300 GPa and 300 K (Fig. 2). The averaged atomic positions derived from a short-MD run (1.5 ps) generated indeed structures with two rotating 3-molecules G-layers (Fig. 2a and 2c) and two static hexagonal 3-molecules G-layers (Fig. 2b and 2d). On the contrary, the averaged atomic positions from a longer MD simulation (5 ps) show the presence of rotating 3-molecules rings in all G-layers (Fig. 2e-2h). Eventually, the averaged atomic positions from the trajectory of longest MD run (13 ps) gave featureless structure due to the extensive diffusive hydrogen motions in G-layers (Fig. 2i-2l) as expected.

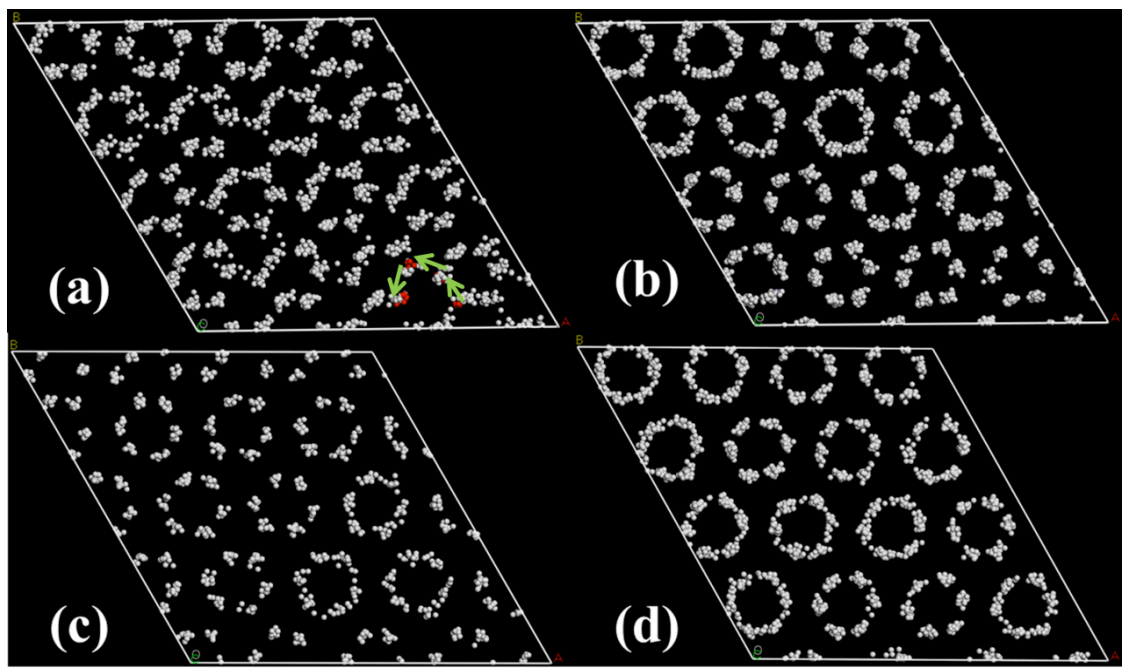

**Fig. 3**: The collected trajectories of long-run MD simulations from the last 7.5 ps (a), 2.5-4.5 ps (b), 6-7 ps (c) and 7.5-8.5 ps (d), respectively. (b)-(d) shows the G-layer reconstructed with time.

Fig. 3a shows the trajectory of the G-layer derived from the last 7.5 ps run of a MD simulation using 768 atoms at 300 K and 300 GPa. It failed to show the 3-molecule ring character. In contrast, the short time trajectory did reveal the molecular nature, as shown in the Fig 3b-3d, where the G-layers reconstructed with time. The analysis presented here clearly shows the occurrence of diffusive hydrogen motions. It is inappropriate to characterize the structure of phase IV from the averaged atomic positions via short-time MD running slices.[15]

To further investigate the proton motion in G-layers, a randomly chosen hydrogen atom, marked by red color, was traced for 10 ps. As shown in Fig. 3a, it is evident that this marked hydrogen atom has migrated from one site to another site in the G-layer (or layer II). A supporting movie of the atomic trajectories of the MD simulation has been prepared and deposited as a supplementary material.[26] The present study shows unambiguously that diffusive hydrogen motion is ubiquitous and an intrinsic property of phase IV in solid hydrogen. Quantum effects will only enhance the diffusive hydrogen motions in the G-layer as have been demonstrated in our previous work.[14]

## Conclusion

The influence of system size effects on diffusive hydrogen motions in phase IV of solid hydrogen at 300 GPa and 300 K has been extensively examined by first-principles MD simulations with various simulation cells of 96-768 atoms. It is demonstrated that long simulation time is essential to reveal the diffusive events. The present results thus confirmed previous MD results.[7, 14] It is concluded that long-range diffusive proton motion in phase IV of solid hydrogen is robust. The present study highlights the importance of time span in MD simulation. It is inappropriate to make arguments on the analysis of MD results, which are far from ergodic.

## Acknowledge

The authors acknowledge the funding supports from China 973 Program under Grant No. 2011CB808200, National Natural Science Foundation of China under Grant Nos. 11274136, 11025418 and 91022029, 2012 Changjiang Scholar of Ministry of Education, and Changjiang Scholar and Innovative Research Team in University (No. IRT1132). Part of calculations was performed in high performance computing center of Jilin University.

*Author to whom correspondence should be addressed: mym@jlu.edu.cn